# Microalgal swimming signatures as indicators of neutral lipids production across growth phases

Jiaqi You[1,2,*], Kevin Mallery[1,3], Douglas Mashek[4], Mark Sanders[5], Jiarong Hong[1,3] and Miki Hondzo[1,2]

[1]St. Anthony Falls Laboratory, University of Minnesota, Minneapolis, MN 55414, USA

[2]Department of Civil, Environmental and Geo- Engineering, University of Minnesota, Minneapolis, MN 55455, USA

[3]Department of Mechanical Engineering, University of Minnesota, Minneapolis, MN 55455, USA

[4]Department of Biochemistry, Molecular Biology and Biophysics, University of Minnesota, Minneapolis, MN 55455, USA

[5]University Imaging Center, University of Minnesota, Minneapolis, MN 55455, USA

*Corresponding author: youxx137@umn.edu

**Grant numbers**:

Legislative-Citizen Commission on Minnesota Resources (LCCMR), Environment and Natural Resources Trust Fund 2015-2016, ID:038-B;

National Science Foundation National Robotics Initiative award, NSF-IIS-1427014.

# Abstract

Microalgae have been shown as a potential bioresource for food, biofuel, and pharmaceutical products. During the growth phases with corresponding environmental conditions, microalgae accumulate different amounts of various metabolites. We quantified the lipid accumulation and analyzed swimming signatures (speed and trajectories) of the motile green alga, *Dunaliella primolecta*, during the lag-exponential-stationary growth cycle at different nutrient concentrations. We discovered significant changes in neutral lipid content and swimming signatures of microalgae across growth phases. The timing of the maximum swimming speed coincided with the maximum lipid content and both maxima occurred under nutrient stress at the stationary growth phase. Furthermore, the swimming trajectories of microalgae signal the growth phase and corresponding intracellular lipid accumulation. Our results provide the potential exploitation of microalgal swimming signatures as indicators for the cultivation conditions and the timing of microalgal harvest to maximize the lipid yield for biofuel production. The findings can also be implemented to explore the production of food and antibiotics from other microalgal metabolites with low energy cost.

**Keywords**: microalgae, lipids, swimming, biofuel

# 1. Introduction

Microalgae have been advocated as a promising source of food, energy and pharmaceutical products. Owing to their high growth rate, competitive lipid production, low environmental impact, and tolerance of harsh environments including wastewater streams, microalgae have been cultivated in large outdoor ponds and closed photo-bioreactors for biofuel production (Chisti, 2007; Griffiths & Harrison, 2009; Rodolfi et al., 2009; Georgianna & Mayfield, 2012;Staples, Malina, & Barrett, 2017; Elrayies, 2018). Microalgae capture solar energy and store it as neutral lipids, primarily comprising of triacylglycerols (TAGs), which can be trans-esterified into biofuel (Guschina & Harwood, 2006; Scott et al., 2010; Chen et al., 2011). Challenges remain in commercializing biofuel production from microalgal lipids due to the high costs and low energy efficiency of large-scale algal cultivation and harvest (Lardon, Hélias, Sialve, Steyer, & Bernard, 2009; Mata, Martins, & Caetano, 2010; Acién, Fernández, Magán, & Molina, 2012; Slade & Bauen, 2013; Vuppaladadiyam, Prinsen, Raheem, Luque, & Zhao, 2018). One key remaining question is when during the cultivation process the biomass yield and intracellular lipid accumulation are maximized. Preferably, the high-lipid timing should be used to initiate harvesting.

Researchers have investigated the microalgal lipid accumulation in response to environmental stresses induced by nutrient limitation. Nitrogen starvation has been reported to facilitate lipid accumulation in green algae cells (Phadwal & Singh, 2003; Chen et al., 2011; Gao, Yang, & Wang, 2013; Mai et al., 2017; Wang, Lai, Karam, de los Reyes, & Ducoste, 2019). The reported impact of nitrogen starvation on algal lipid production differs among species (Griffiths & Harrison, 2009) and is also affected by experimental design and the extent of nitrogen starvation. Environmental stresses also influence the motility of microalgal cells. Swimming behaviors of green alga *Dunaliella* are influenced by fluid flow conditions (Chengala, Hondzo, Troolin, & Lefebvre, 2010; Chengala, Hondzo, & Mashek, 2013), and increased swimming velocities of freshwater green alga *Chlamydomonas* have been observed under nitrogen limitation (Hansen, Hondzo, Mashek, Mashek, & Lefebvre, 2013). Limited research has been done to directly relate swimming behaviors of algal cells to their metabolism (e.g., lipid accumulation) and physiological stages.

Due to their high lipid content, halotolerance and lack of rigid cell wall, the motile green algae, *Dunaliella*, was chosen as the model organism to investigate swimming signatures and intracellular lipid accumulation under different environmental conditions. Our primary hypothesis is that swimming velocities and trajectory patterns of *Dunaliella primolecta* (*D. primolecta*) signal the intracellular lipid accumulation. Micro-particle tracking velocimetry (micro-PTV) and digital inline holographic particle tracking velocimetry (DIH-PTV) were employed to analyze the swimming signatures (velocities and trajectories) of *D. primolecta* cells with different intracellular lipid content. The results demonstrate statistically different swimming signatures of *D. primolecta* during the lag, exponential, and stationary growth phases, which correspond to different lipid content in the cells. Since microalgal signatures can be obtained *in situ* and in real-time, the findings will be instrumental for cultivating and harvesting microalgal cells with maximum lipid content.

## 2. Materials and methods

### 2.1. Strain selection and cultivation

*D. primolecta* is a species of unicellular, motile, bi-flagellated green algae which lives in brackish water, making it preferable organism for monoculture cultivation under field conditions. The lipid content in *Dunaliella* ranges from 15% - 40%, and most of the TAGs are accumulated in cytoplasmic lipid droplets (Zweytick, Athenstaedt, & Daum, 2000; Huesemann & Benemann, 2009; Davidi, Katz, & Pick, 2012). Batch cultures of *D. primolecta* UTEX LB1000 were incubated in the three groups of modified Erdschreiber's medium with 30%, 70% and 100% initial nitrogen concentrations and labeled as 30%N, 70%N and 100%N, respectively (Supplementary, SM, S1.1). All the cultures were started at approximately $2\times10^4$ cells/mL in 250 mL Erlenmeyer flasks, and incubated in an incubator at 20°C under PAR of 51 $\mu E/m^2 s$ with 14/10-hour light/dark cycles, maintained mixed by a shaker table (New

Brunswick Scientific Co., USA) at 30 rpm and manually shaken twice daily. Trial cultivation experiments with above conditions had been conducted twice to have a clear understanding of the cultivation duration before the cultivation of *D. primolecta* for lipid and swimming analyses. The cultivation lasted for 51 days, including the entire growth cycle until the beginning of the decline phase.

## 2.2. Nitrate analysis, dry weight measurement, and microscopic observation

Samples were collected from replicate *D. primolecta* cultures with 30%N, 70%N and 100%N growth medium, respectively, at each growth phase for the measurement of dry cell weight (SM, S1.2) and nitrate concentration. Nitrate analysis was performed with the Trilogy laboratory fluorometer (Turner Designs Inc., USA) and the nitrate test kit (LaMotte Nitrate Nitrogen Kit, 0.25 -10 ppm, LaMotte company, USA). *D. primolecta* cells from samples of lag, exponential, early stationary and late stationary phases were observed under a Nikon Eclipse 90i microscope with a 100X oil immersion DIC objective and images were captured using a Nikon DS-Fi2 color CCD camera in the field of 1260 x 960 pixels at 0.07 µm/pixel (SM, S1.3).

## 2.3. Determination of intracellular neutral lipid content

The relative quantity of intracellular neutral lipid of *D. primolecta* cells was determined by the Nile red fluorescence according to Chen et al. (2009) throughout the growth cycle. Samples were taken from replicate cultures of each growth medium group (i.e., 30%N, 70%N and 100%N) on each measurement day. *D. primolecta* samples were incubated in 96-microwell plates in the dark at 25˚C for 30 min to get the algal cells stained. The fluorescence emissions were then recorded by a BioTek Synergy Neo2 Hybrid multi-mode reader (BioTek Instruments, Inc., USA), with an excitation wavelength of 530 nm and an emission wavelength of 575 nm. The fluorescence signals were normalized by the cell concentration to the fluorescence intensity per $10^7$ cells. The neutral lipid content of algal samples from eight different cultures at various growth phases were determined by the gravimetric method modified from Bligh and Dyer (1959) and Alonzo and Mayzaud (1999) to calibrate the Nile red fluorescence intensity with the neutral lipid content (SM, S1.4).

## 2.4. Measurement of swimming velocities

Swimming velocities of *D. primolecta* were measured at lag, exponential, early stationary and late stationary phases. A micro-PTV system developed by TSI Inc. (USA) was employed to quantify the swimming velocities. The micro-PTV system set up was similar to the one described in Chengala et al. (2010) with modifications (Figure S1). At least four samples were collected from each culture group (30%N, 70%N and 100%N) and two image sequences were recorded at a frame rate of 5 Hz ($\Delta T=200000$ μs) to provide at least 100 pairs of frames for each sample. The processing of images was based on PTV (particle tracking velocimetry) analysis, which measured the horizontal velocities (x-direction and y-direction) of each particle (algal cell) in pairs of images by tracking the displacement of particles between each pair of frames and generating velocity vectors (SM, S1.5).

## 2.5. Tracking of swimming trajectories

The swimming trajectories of *D. primolecta* were recorded using digital inline holography (DIH). DIH uses a single camera to record the pattern generated from the interference between the scattered light from objects and the unscattered portion of the coherent light for illumination, providing a low cost and compact solution for 3D imaging of biological samples (Xu et al. 2001). Compared to conventional microscopy methods, DIH-PTV has a much larger depth of focus and is capable of tracking objects in 3D (Yu, Hong, Liu, & Kim, 2014). DIH has been recently applied as particle tracking velocimetry (PTV) technique to study fluid motion and the biolocomotion of biological organism (Sheng et al. 2007, Kumar et al. 2016, Toloui et al. 2017, You et al. 2017). Our DIH setup consists of a 532 nm diode laser (Thorlabs CPS532), an optical spatial filter and collimating lens assembly, 5X microscopic objective (Mitutoyo 10X/0.14 NA), and a CCD camera (Flare 2M360-CL) (Figure S2). The holograms were recorded at a framerate of 100 Hz with an exposure time of 50 μs and cropped to a 1024 pixels × 1024 pixels window for processing. Measurements of replicate 30%N cultures were performed at the lag, exponential, and early stationary phases. Recordings of five samples were made for each culture with recording volume of 1 μL. Each sample was recorded for 20 seconds (2000 frames) with the exception of the lag stage samples which were recorded for 60 seconds in order to increase the total number of cells recorded at the lowest concentration. Positions of each cell in the recorded hologram were extracted using the Regularized Inverse Holographic Volume Reconstruction (RIHVR) method presented by Mallery et al. (2019). The cell positions were tracked in time using the method of Crocker and Grier (1996) to produce trajectories illustrating the swimming pattern of each cell.

The trajectories were classified into five swimming patterns: 1) circular which rotate about a fixed point with minimal net motion; 2) helical which rotate about a moving point, forming a helix; 3) random walk which do not rotate and have minimal directionality; 4) linear which have no rotation and minimal deviation from a principle direction of motion; 5) meandering which is similar to linear but with greater deviation from the principle direction. The trajectories were automatically labeled using a binary decision tree based on a set of measured trajectory features.

### 2.6. Statistical analyses

One-way ANOVA tests were conducted for each culture group to determine the significance of differences in the neutral lipid content and swimming speed of *D. primolecta* cells at different growth phases ($df = 3$, $\alpha = 0.05$). One-way ANOVA tests were also conducted at each growth phase to determine the significance of differences in the neutral lipid content and swimming speed of *D. primolecta* cells in cultures with different initial nitrogen availability ($df = 2$, $\alpha = 0.05$). To determine the significance of differences in the swimming behavior among growth phases (the lag, exponential and early stationary phases), the one-way ANOVA tests were conducted for the swimming speed and fine motion frequency from the DIH-PTV results ($df = 2$, $\alpha = 0.05$).

## 3. Results and Discussion

### 3.1. Intracellular neutral lipid content accumulation

During the growth cycles, all the cultures maintained at the lag phase for the first few days adapting to the new environment, and then grew exponentially, after which the cells entered the stationary phase with constant populations until the cell population started to decline (Figure 1a). The growth cycle can be divided into four growth phases: 1) lag phase - day 1, i.e. 24 hours after the beginning of cultivation; 2) exponential phase - day 16; 3) early stationary phase – 2 or 3 days after the beginning of the stationary phase (day 25 for 30%N cultures and day 35 for 70%N and 100%N cultures); and 4) late stationary phase - the stage right before the decline phase (day 45).

The intracellular neutral lipid content of *D. primolecta* changed correspondingly to the growth phases (Figure 1b). The neutral lipid content slightly increased but remained relatively low for all cultures from lag to exponential growth phases. Starting from the late exponential phase, the neutral lipid content of the 30%N cultures increased drastically, reaching the peak at the early stationary phase (day 25) and decreasing at the late stationary phase. The increase of intracellular neutral lipid content in the 70%N and 100%N cultures were not drastic until the late stationary phase. The peaks of neutral lipid level occurred on day 45 for 70%N and 100%N cultures, and the peaks were lower than the peak for the 30%N cultures. Converting the Nile red fluorescence intensity to neutral lipid content (standard curve in Figure S3), the intracellular neutral lipid content for all the cultures was under 5% of cell dry weight during lag and exponential phases. The neutral lipid content of cells in the 30%N cultures reached its peak of 35.9% of dry weight at the early stationary phase and decreased to 7.7% of dry weight at the late stationary phase. In contrast, the intracellular neutral lipid content in 70%N and 100%N cultures remained under 10% of dry weight and increased to the peak of 20.5% and 16% of dry weight, respectively, at late stationary phase. Both the timing and the extent of the peak of the intracellular neutral lipid content were negatively influenced by the initial nitrogen availability in the cultures.

Relating the neutral lipid content with the nitrate concentration, we observed that the sharp rise of the neutral lipid content did not occur until the nitrate concentration in the culture became less than 1 mg/L (Figure S4). The nitrate concentration in the 30%N cultures decreased below this threshold during the late exponential phase, around day 17, soon after which the neutral lipid level reached the peak. The nitrate concentration of the 70%N and 100%N cultures decreased from slightly above the threshold to below the threshold during the stationary phase, day 35 to day 37, and therefore showed the peaks of neutral lipid at late stationary phase. For all the three groups of cultures, the neutral lipid content was approximately 0.1 mg per $10^7$ cells when the nitrate concentration was near the 1 mg/L. Our experimental results agree with the previous research by Gao et al. (2013), Mai et al. (2017) and Wang et al. (2019) indicating that *Dunaliella* cells respond to the nitrogen starvation by rapidly transferring previously stored carbon to neutral lipids and enhancing lipid accumulation. The timing of peak values is also consistent with that reported in Chen et al. (2011) which stated that the lipid content of *Dunaliella* cells reaches the highest value at the stationary phase in both the normal medium culture and the N-deficient medium culture.

Under the light microscope, changes in the intracellular structure and contents of *D. primolecta* cells were observed at the four growth phases (Figure 2). The cup-shaped chloroplast with pyrenoid was observed in cells at lag and exponential phases (Figure 2a-c, d-f). During the stationary phase, the cytoplasm of cells was no longer as clear, and refractive starch granules appeared in the chloroplast. Starting from the early stationary phase, the cells in the 30%N cultures depicted an intracellular composition with more closely packed organelles and some neutral lipid globules which were not present in cells at the lag and exponential phases (Figure 2g). Most of the cells at the late stationary phase were observed to be spherical or near-spherical due to the unfavorable growth conditions and exhibited a yellow-brownish color associated with the age of cells and the increase of intracellular neutral lipid (Figure 2j-l). The microscopic observations were consistent with the study by Eyden (1975) of the *Dunaliella primolecta* cells under light and electron microscope, and the study by Davidi et al. (2012), which reported observation of a fragmented chloroplast with multiple cytoplasmic lipid droplets in the N-deficient cells. It also agrees with the report by Chen et al. (2011) that intracellular neutral lipid bodies started to be observed by the third day of nitrogen deprivation, and the statement in Hu et al.(2008) that the deposition of TAGs into cytosolic lipid bodies play active roles in stress response.

## 3.2. Microalgal swimming speed and trajectories

The motility of *D. primolecta* changed corresponding to growth phases and the initial nitrogen availability in the growth media. The scatter plots of horizontal velocities in *x*- and *y*- directions were all in circular/near-circular patterns, centered around the origin (0,0) and expanding or shrinking showing the velocity changes with growth phases (Figure S5). All the velocities from the experimental results were within the range of -120 to 120 µm/s. The velocities were relatively low at lag phase and increased from lag to early stationary phase for all culture groups. The velocities of the cells in the 30%N cultures at the early stationary phase concentrated at higher values with a decrease when reaching the late stationary phase, while the velocities of the cells in the 70%N and 100%N cultures kept increasing till the late stationary phase.

The swimming speed, *v*, of *D. primolecta* cells in the horizontal *x*-*y* plane were all within 5 to 120 µm/s. The distribution of the swimming speed also varied with growth phases and initial nitrogen availabilities (Figure 3). The probability density function (PDF) of swimming speed for each culture group at each growth phase was approximated by a Kernel distribution, and the approximations were confirmed by the Chi-square goodness-of-fit tests at $\alpha = 0.05$. The variation of swimming speed distribution indicated that the swimming speed of *D. primolecta* cells increased from the beginning of a growth cycle until reaching a peak during the stationary (early stationary or late stationary) phase, when the nutrient deprivation occurred and the cell population stopped increasing. The 30%N cultures showed left-skewed speed distribution earlier than the 70%N and 100%N cultures, and the mean of the distribution for the early stationary 30%N cultures (81 µm/s) was higher than that for the late stationary 70%N and 100%N cultures (67 µm/s and 62 µm/s, respectively), which implied an impact of the initial nitrogen availability on the variation of swimming speed.

Measurements of the *D. primolecta* swimming velocities using DIH-PTV showed similar trends to the micro-PTV measurements. The lag phase showed the lowest mean swimming speed, with increasing speeds seen in both the exponential and early stationary stages. There was no significant difference in speed among the three nitrogen levels for the lag or exponential phase. Swimming trajectory measurements using DIH-PTV indicate a change in the complex behavior modes of the swimming cells (Figure 4a-c). We observed five swimming patterns including circular, helical, random walk, meandering, and linear. These can be further grouped into fine motions (circular and helical) and gross motions (linear and meandering) with random walk serving as an intermediate that is neither fine nor gross. We trained a binary decision tree to autonomously classify the measured trajectories into one of these classes. We applied this classification to $n = 421$ trajectories from the lag phase, $n = 6306$ from the exponential phase, and $n = 7874$ from the early stationary phase. The most significant trend is the decrease in the frequency of fine motions from exponential to early stationary phase (Figure 4d-f). The helical and circular behaviors are hardly seen in the early stationary stage. The fine motions account for 48% of tracks in the lag phase, 16% in the exponential phase, and only 2% in the early stationary phase.

### 3.3. Indication of neutral lipid accumulation by swimming signatures

The experimental results revealed that both the intracellular neutral lipid accumulation and the swimming signatures of *D. primolecta* cells changed under the stress of nutrient limitation. Noticeable similarities in the trend of the changes and timing of the peak values were discovered between the lipid accumulation and swimming speed (Figure 5). The neutral lipid content and swimming speed of cells in all the three culture groups increased from the lag phase to the early stationary phase. Cells in the 30%N cultures achieved their maximum neutral lipid content and swimming speed at the early stationary phase, while the maxima for the 70%N and 100%N cultures were achieved at the late stationary phase. On the other hand, the initial nitrogen availability was observed to have a negative impact on the extent of changes in both neutral lipid and swimming speed. The lower the initial nitrogen concentration in the growth medium, the higher the maximum neutral lipid and swimming speed were. The cells in the 30%N cultures could accumulate neutral lipid content as high as 35.9% of dry weight when their average swimming speed reached the highest value of 81 μm/s, while the peak values of neutral lipid content for the 70%N and 100%N cultures were 20.5% and 16% of dry weight, coinciding with their peak swimming speeds of 67 µm/s and 62 µm/s, respectively.

The statistical results from the one-way analysis of variance (ANOVA) showed the consistency between neutral lipid accumulation and swimming speed variation of *D. primolecta* cells (Table S1). Both the neutral lipid and swimming speed of cells changed significantly ($\alpha = 0.05$, $P < 0.01$) across the four growth phases. No significant differences in the neutral lipid content or the swimming speed were determined among the three cultures with various initial nitrogen availabilities in the lag and exponential phases, while the differences in both the neutral lipid content and the swimming speed were significant among the three culture groups at the early and late stationary phases ($\alpha = 0.05$, $P < 0.01$). The results of the statistical analyses support our hypothesis that unicellular green alga *D. primolecta* signals the intracellular lipid accumulation by swimming speed.

The swimming speed of *D. primolecta* cells and the nitrogen transport to the cells by molecular diffusion were quantified by the Peclet number,

$$Pe = \frac{t_{diff}}{t_{swim}} \qquad (1)$$

where $t_{diff}$ is the diffusion time taken for nitrate to diffuse over an average cell diameter, and $t_{swim}$ is the swimming time taken for *D. primolecta* cells to swim over a distance equivalent to an average cell diameter. The diffusion time and swimming time can be expressed as

$$t_{diff} = \frac{L_c^2}{D_{NO_3^-}} \qquad (2)$$

$$t_{swim} = \frac{L_c}{v} \qquad (3)$$

where $v$ is the average swimming speed of the cells (µm/s), $L_c$ is the average diameter of the cells (µm) obtained from the micro-PTV measurements (data shown in Table S2), and $D_{NO_3^-}$ is the molecular diffusion coefficient of nitrate (1700 μm²/s). Substituting equation (2) and (3) into equation (1), the Peclet number can be expressed as

$$Pe = \frac{v L_c}{D_{NO_3^-}} \tag{4}$$

The highest Peclet number, $Pe = 0.5$, in this experiment was achieved by the 30%N cultures at the early stationary phase, accompanied by the highest intracellular neutral lipid content. Two trends were observed between the neutral lipid content and the Peclet number (Figure 6). A linear trend from lag to exponential phase ($R^2 = 0.97$, $0.2 < Pe < 0.31$) and another linear trend with a much steeper slope for exponential, early stationary and late stationary phases ($R^2 = 0.94$, $0.31 < Pe < 0.5$) were observed. The results indicate a transition point occurring around the late exponential phase. In the lag and exponential phases, the nitrogen availability was sufficient for the cells to grow, so the change in swimming speed was not triggered by the nutrient deficiency, and therefore the neutral lipid content only increased slightly with the increase of Peclet number. In contrast, once the cells entered the late exponential phase (i.e., after the time point of the maximum growth rate), the nitrogen availability in the environment shifted from sufficient to limited, which triggered the increase of the swimming speed of cells (greater Peclet number) to actively seek the nutrient while at the same time increasing the neutral lipid accumulation in the cells. This linear correlation corresponding to the late exponential – stationary phases provides a way to reflect the neutral lipid accumulation by measuring the swimming speed of cells under the conditions of nutrient limitation and nutrient deprivation.

The changes of swimming patterns identified from the *D. primolecta* swimming trajectories also correspond to the intracellular neutral lipid content changes across growth phases. The statistical one-way ANOVA results indicate that the percentage of cells exhibiting fine motions are statistically different at each growth phase ($\alpha = 0.05$, $P < 0.01$) (Table S3). The variation of swimming behaviors across growth phases could be associated with the nutrient availability and chemotaxis of microalgal cells. The nutrient distribution in the growth environment is hardly homogeneous – nutrient patchiness forms for many reasons. Chemotactic response of *Dunaliella* cells to the ammonium and several amino acids have been reported (Sjoblad, Chet, & Mitchell, 1978). Akin to the run-tumble transitions of bacterial chemotaxis (Adler, 1966), the cells move towards a nutrient patch through the gross (linear and meandering) motions while remaining in the patch to utilize the nutrients through the fine (helical and circular) motions. At early growth phases, the nutrients are sufficient, so the frequency of fine motions is high since the shorter net displacement of helical and circular motions allow the cells to maintain in a nutrient-rich region for longer time and maximize the nutrient uptake. When the photosynthesis is limited by the low nutrient, the algal cells tend to become positively chemotactic which makes them travel faster in linear/meandering motions towards certain attractants (Lee, Lewitus, & Zimmer, 1999). At the stationary growth phase, the fine motions are significantly reduced due to nutrient scarcity and most of the cells exhibit larger net displacement with the gross motions to explore potential nutrient sources.

The reduction of the fine motions corresponds to the accumulation of intracellular neutral lipid content as both are induced by the nutrient limitation. When the intracellular neutral lipid content reached the maximum at the early stationary phase, the fine motions were hardly seen. There were approximately 8-fold differences in both the fine motions and neutral lipid content between the exponential and early stationary phases for the 30%N cultures. The strength of transition suggests that the relative frequency of behavior modes could be a strong indicator for the growth phase, and by association the neutral lipid content.

### 3.4. Application of swimming signatures to biofuel production

Our findings of the association between microalgal swimming signatures and intracellular neutral lipid content inspire a low-cost scalable sensor system utilizing DIH-PTV for high-throughput monitoring of the lipid content in industrial-scale biofuel production. Such a system would improve control of cultivation conditions and harvest timing of microalgae, leading to a substantial biofuel yield and low-cost algae-based biofuels. Conventional methods for intracellular neutral lipid determination, such as lipid separation by column or thin-layer chromatography, fatty acid profiling by gas chromatography, and fluorometric determination of neutral lipids by Nile red / BODIPY (Cooksey, Guckert, Williams, & Callis, 1987; Elsey, Jameson, Raleigh, & Cooney, 2007; Chen et al., 2009; Siaut et al., 2011; Govender, Ramanna, Rawat, & Bux, 2012; Moheimani, Borowitzka, Isdepsky, & Sing, 2013; Talebi et al., 2015; Yang, He, & Hu, 2015), either require tedious extraction and/or transesterification work and expensive equipment, or test only small volume samples which are less representative for large-scale cultivations. In contrast to current lipid measurement methods, DIH-PTV has much higher throughput and better scalability. DIH sensors are low cost, small, and easily integrated with other instruments so that a network of DIH sensors could monitor multiple sites efficiently in open ponds or closed photo-bioreactors for industrial-scale biofuel production.

Exploiting microalgal swimming signatures as an indicator in microalgal growth and lipid accumulation monitoring will potentially influence the pipeline of microalgal biodiesel production by enabling more optimal harvest timing and control of cultivation conditions (Figure S6). Our results show that the intracellular neutral lipid content can increase 3.6 folds in only three days. The peak lipid accumulation at the early stationary phase is 20 folds higher than that at the exponential phase and 4.7 folds higher than that at the late stationary phase. Since lipid production is very sensitive to harvest timing, utilizing the drastic decrease of fine motions and increase in swimming speed as an indicator of maximum lipid accumulation will enable more optimal harvest timing. This timing will also reduce the cost of consumables (including water, nutrient supplies, and electricity), some of the most substantial costs in biofuel production (Barsanti & Gualtieri, 2018). This can lower the net cost per unit biodiesel production, which is a major bottleneck inhibiting commercialization (Ruiz et al., 2016; Barsanti & Gualtieri, 2018).

These factors will together decrease the net energy ratio (NER, the energy input in cultivation and refinement divided by the energy output from the biomass), which is an indicator of environmental sustainability and the overall energetic effectiveness of the biodiesel production (Slade & Bauen, 2013). Achieving a lower NER will make the microalgal biodiesel production more economically competitive. Besides reducing the net cost per unit biodiesel, the higher biomass and lipid yield can also result in increased beneficial by-products from the lipid extraction and transesterification processes. Appropriate use of the remaining biomass for animal feed, ethanol, or biogas, and glycerol for food industry, pharmaceutical applications, or personal care will contribute to the commercial viability of microalgal biofuel production (Scott et al., 2010; Doshi, Pascoe, Coglan, & Rainey, 2016; Gerber, Tester, Beal, Huntley, & Sills, 2016). Overall, exploiting swimming signatures in a low-cost scalable sensor system for inline high-throughput monitoring of microalgal growth and lipid accumulation will potentially lead the microalgae-based biofuel production to a more cost-effective and energy-effective future.

## Acknowledgements

The work was supported by the Legislative-Citizen Commission on Minnesota Resources (LCCMR), Environment and Natural Resources Trust Fund 2015-2016 [Assessing the Increasing Harmful Algal Blooms in Minnesota Lakes, ID:038-B]; and the National Science Foundation National Robotics Initiative award [NSF-IIS-1427014]. The cell imaging was done using the Nikon Eclipse 90i microscope with the assistance of Guillermo Marques and John Oja at the University of Minnesota – University Imaging Centers. The authors would also like to thank Wenqi Cui and Enxiang Zhang for the technical support in neutral lipid determination, Dr. Dan Troolin for technical support in micro-PTV measurements and Jaewoo Jeong for part of the holography image processing.

# Figure Legends

**Figure 1**. Growth curves and intracellular neutral lipid content of *D. primolecta* cultures. (**a**) Cell concentrations of an entire growth cycle of the three groups of *D. primolecta* cultures. Error bars correspond to the s.d. ($n$=8). (**b**) Intracellular neutral lipid content reflected by the intensity of Nile red fluorescence detected at 530/575 nm excitation/emission wavelength for 30%N cultures (□), 70%N cultures (○) and 100%N cultures (Δ). Error bars correspond to the s.d. ($n$=4).

**Figure 2.** *D. primolecta* cells at various growth phases under a light microscope. (**a**)-(**c**) Cells at the lag phase of 30%N, 70%N and 100%N cultures, respectively. (**d**)-(**f**) Cells at the exponential phase of 30%N, 70%N and 100%N cultures, respectively. (**g**)-(**i**), Cells at the early stationary phase of 30%N, 70%N and 100%N cultures, respectively. (**j**)-(**l**) Cells at the late stationary phase of 30%N, 70%N and 100%N cultures, respectively.

**Figure 3**. Fitted probability density functions for *D. primolecta* swimming speeds at four growth phases. The probability density functions were produced by fitting the data into Kernel distributions (goodness-of-fit confirmed by Chi-square tests at $\alpha$=0.05).

**Figure 4.** Classified swimming trajectories of D. *primolecta* cells in 30%N cultures. (**a**)-(**c**) Selection of $n$=400 random trajectories from each of the three stages measured with DIH-PTV. Trajectories are colored according to their assigned behavior mode. (**d**)-(**f**) The identity frequency for each behavior mode for each of the growth stages.

**Figure 5**. Intracellular neutral lipid content and swimming speed of *D. primolecta* at four growth phases. (**a**) Intracellular neutral lipid content reflected by the intensity of Nile red fluorescence for 30%N, 70%N and 100%N cultures at four growth phases. Error bars correspond to the s.d. ($n$=4). (**b**) The average swimming speed of 30%N, 70%N and 100%N cultures at four growth phases. Error bars correspond to the s.d. (*n*=200 for lag phase, *n*=400 for exponential phase and *n*=600 for early stationary and late stationary phase).

**Figure 6.** Neutral lipid content versus the Peclect number of D. *primolecta* cells. Data points (*n*=12) are from 30%N, 70%N and 100%N cultures at four growth phases. The vertical dash dot line (–·–) separates the graph into two areas based on growth phases: from lag to early exponential phase on the left, and from late exponential to stationary phase on the right. The dot line (··) corresponds to the linear trend line fitted to the data points from the lag and exponential phases (*n*=6, $R^2$=0.97). The dash line (– –) corresponds to the linear trendline fitted to the data points from exponential, early stationary and late stationary phases (*n*=9, $R^2$=0.94).

# Appendix – Supplementary Material

Detailed information of the methods and materials, and supplementary figures and tables regarding the experimental set up and results as noted in text is provided in the document of Supplementary Material.